\documentclass[runningheads,a4paper]{llncs}
\usepackage{color,wrapfig,lipsum,booktabs}
\usepackage{amssymb}
\usepackage{graphicx}
\usepackage{times}
\usepackage{epsfig}
\usepackage{graphicx}
\usepackage{amsmath}
\usepackage{amssymb}
\usepackage{stmaryrd}
\usepackage{adjustbox} 
\usepackage{bm}
\usepackage{multirow}
\usepackage{enumitem}
\usepackage{balance}
\usepackage{flushend}
\usepackage{booktabs}
\usepackage{graphicx}

\usepackage{url}
\urldef{\mailsa}\path|{lsun29, psyu,caobokai}@uic.edu|
\urldef{\mailsc}\path|csyqwang@comp.polyu.edu.hk|    
\urldef{\mailsb}\path|witty@cse.unl.edu|
\urldef{\mailsd}\path|aleow@psych.uic.edu|

\begin{document}

\mainmatter  

\title{Sequential Keystroke Behavioral Biometrics for Mobile User Identification via Multi-view Deep Learning}


\author{Lichao Sun, Yuqi Wang+, Bokai Cao, \\Philip S. Yu,  Witawas Srisa-an* and Alex D Leow}
%

\institute{University of Illinois at Chicago,
Chicago, IL 60607\\
*University of Nebraska--Lincoln,
Lincoln, NE 68588\\
+Hong Kong Polytechnic University,
Kowloon, Hong Kong \\
\mailsa\\
\mailsd\\
*\mailsb\\
+\mailsc\\
}

%
%

\maketitle

\begin{abstract}

With the rapid growth in smartphone usage, more
organizations begin to focus on providing better
services for mobile users.  User identification can help
these organizations to identify their customers and
then cater services that have been customized for them.
Currently, the use of cookies is the most common form
to identify users. However, cookies are not easily
transportable (e.g., when a user uses a different login
account, cookies do not follow the user). This
limitation motivates the need to use behavior biometric
for user identification.  In this paper, we propose
\textsc{DeepService}, a new technique that can identify
mobile users based on user's keystroke information
captured by a special keyboard or web browser. Our evaluation results
indicate that \textsc{DeepService} is highly accurate
in identifying mobile users (over 93\% accuracy).  The
technique is also efficient and only takes less than 1 ms to
perform identification.

\end{abstract}

\section{Introduction}
Smart mobile devices are now an integral part of daily
life; they are our main interface to cyber-world. We
use them for on-line shopping, education,
entertainment, and financial transactions.
As such, it is not surprising that companies are
working hard to improve their mobile services to gain
competitive advantages. Accurately and non-intrusively
identifying users across applications and devices is one
of the building blocks for better mobile experiences,
since not only companies can attract users based on
their characteristics from various perspectives, but
also users can enjoy the personalized services without
much effort~\cite{scikit-learn}.

User identification is a fundamental, but yet an open
problem in mobile computing. Traditional approaches
resort to user account information or browsing history.
However, such information can pose security and privacy
risks, and it is not robust as can be easily changed, e.g., the user
changes to a new device or using a different
application. Monitoring biometric information including
a user's typing behaviors tends to produce consistent
results over time while being less disruptive to 
user's experience. Furthermore, there are different
kinds of sensors on mobile devices, meaning rich
biometric information of users can be simultaneously
collected.  Thus, monitoring biometric information
appears to be quite promising for mobile user
identification. 

To date, only a few studies have utilized
biometric information for mobile user identification 
on web browser~\cite{abramson2013associative,zhang2014mfingerprint}.
Important questions such as what kind of biometric
information can be used, how does one capture user
characteristics from the biometric, and what accuracy of
the mobile user identification can be achieved, are
largely unexplored. Although there are some researches
on mobile user authentication through biometrics~\cite{feng2012continuous,zhao2013continuous},
authentication is a simplified version of
identification, and directly employing authentication
would either be infeasible or lead to low accuracy. This work
focuses on mobile user identification, and could also
be applied to authentication.


In this paper, we collect information from basic
keystroke and the accelerometer on the phone, and then
propose \textsc{DeepService}, a multi-view deep learning
method, to utilize this information. To the best of our
knowledge, this is the first time multi-view deep
learning is applied to mobile user identification.
Through several empirical experiments, we showed that
the proposed method is able to capture the user
characteristics and identify users with high accuracy.

Our contributions are summarized as follows.
\begin{enumerate}
  \item We propose \textsc{DeepService}, a multi-view deep learning method, to utilize easy to collect user biometrics for accurate user identification.
 \item We conduct several experiments to demonstrate the effectiveness and superiority of the proposed method against various baseline methods.
  \item We give several analyses and insights through the experiments.
\end{enumerate}

The rest of this paper is organized as follows. Section~\ref{relatedwork} provides background information on deep learning, and reviews prior research efforts related to this work. Section~\ref{overview} introduces \textsc{DeepService} and describes the design and implementation details. Section~\ref{experiment} reports the results of our empirical evaluation on the performance of \textsc{DeepService} with respect to other learning techniques. The last section concludes this work and discusses future work.

\section{Background and Related Work} \label{relatedwork}

In this section, we provide additional background
information on deep learning structure, and review prior
research efforts related to our proposed work.

\subsection{Background on Deep Learning Structure}

Deep learning is a branch of machine learning based
on a set of algorithms that attempt to model high level
abstractions in data, which is also called deep structured
learning, deep neural network learning or deep machine
learning. Deep learning is a concept and a framework
instead of a particular method.  There are two main
branches in deep learning: Recurrent Neural Network
(RNN) and Convolutional Neural Networks (CNN). CNN is
frequently used in computer vision areas and RNN is
applied to solve sequential problems such as nature
language process. The simplest form of an RNN is shown as follows:
\begin{equation*}
    h_k =\phi (Wx_k + Uh_{k-1})
\end{equation*}
where $h_k$ is the hidden state, $W$ and $U$ are parameters need
be learned, and $\phi (.)$ is the is a nonlinear transformation
function such as tanh or sigmoid.




Long Short Term Memory network (LSTM) is a special case
of RNN, capable of learning long-term
dependencies~\cite{hochreiter1997long}. Specifically, RNN only
captures the relationship between recent keystroke
information and uses it for prediction. LSTM, on the
other hand, can capture long-term dependencies.
Consider trying to predict the tapping information in
the following text ``I plan to visit China ... I need
find a place to get some Chinese currency''. The word
``Chinese'' is relevant with respect to the word
``China'', but the distance between these two words is
long. To capture information of the long-term dependencies,
we need to use LSTM  instead of the standard RNN
model.

While LSTM can be effective, it is a complex deep
learning structure that can result in high overhead.
Gated Recurrent Unit (GRU) is a special case of LSTM
but with simpler structures (e.g., using less
parameters)~\cite{chung2014empirical}. In many problem
domains including ours, GRU can produce similar results
to LSTM. In some cases, it can even produce better
results than LSTM.  In this work, we implemented Gated
Recurrent Unit (GRU). 

Also note that with GRU, it is quite straightforward
for each unit to remember the existence of a specific
pattern in the input stream over a long series of time
steps comparing to LSTM. Any information and patterns
will be overwritten by update gate due to its
importance.

We can build single-view single-task deep learning
model by using GRU as shown in Figure~\ref{fig:mvmtl}(b).
We choose any one view of the dataset such as the view of
alphabet as used in this study. We use the normalized
dataset as the input of GRU. GRU will produce a final
output vector which can help us to do user
Identification. A typical GRU is formulated as:
\begin{align*}
z_t &= \sigma_g (W_zx_t + U_zh_{t-1}) \\ 
r_t &= \sigma_g(W_rx_t + U_rh_{t-1})\\
\tilde{h}_t &= tanh (Wx_t + U(r_t \odot h_{t-1})) \\
h_t &= z_t\tilde{h}_t + (1- z_t)h_{t-1} \\
\end{align*}
\noindent
where $\odot$ is an element-wise multiplication. $\sigma_g$ is the sigmoid and equals  $1/(1+e^{-x})$. $z_t$ is the update gate which decides how much the unit updates its activation or content. $r_t$ is reset gate of GRU, allowing it to forget the previously computed state.

In Section~\ref{overview}, we extend the single-view
technique to develop \textsc{DeepService}, a multi-view
multi-class framework. 

\subsection{Related Work}

Most previous works focus on user authorization, 
rather than identification, based on biometrics. For example,
there are multiple approaches to get physiological
information that include facial features and iris
features~\cite{goh2003computation,de2001iris}.
This physiological information can also be used for
identification, but it requires extra permission from
the users. Our method uses behavioral biometrics to
identify the users without any cookies or other
personal information.

Recently, more research work on continuous
authorization  problem has emerged for mobile users.
Some prior efforts also focus on studying touchscreen
gestures~\cite{zhao2013continuous}
or behavioral biometric behaviors such as reading,
walking, driving and
tapping~\cite{bo2014continuous,bo2013you}. There are
research efforts focusing on offering better security
services for mobile users. However, their security
models have to be installed on the mobile devices. They
then perform binary classifications to detect
unauthorized behaviors. Our work focuses on building a
general user identification model, which can also be
deployed on the web, local devices or even network
routers. Our work also focuses on improving users'
experience through customized services including
providing recommendations and relevant advertisements.

Recently, some research groups focus on mobile user
identification based on web browsing
information~\cite{abramson2013associative,zhang2014mfingerprint}. 
Abramson and Gore try to
identify users' web browsing behaviors by analyzing
cookies information. However, our model has been
designed to target harder problems without using trail
of information such as cookies or browsing history. We,
instead, use behavioral biometrics to identify users.
Information needed can be easily collected from web
browser using Javascript. 


\section{\textsc{DeepService}: A Multi-view Multi-Class Framework for User Identification on Mobile Devices} \label{overview}

\textsc{DeepService} is a multi-view and multi-class
identification framework via a deep structure.  It
contains three main steps to identify each user from
several users. This process is shown in
Figure~\ref{fig:framework} and summarized below:

\begin{figure*}[tbh!]
    \centering
    \includegraphics[width=0.9\textwidth]{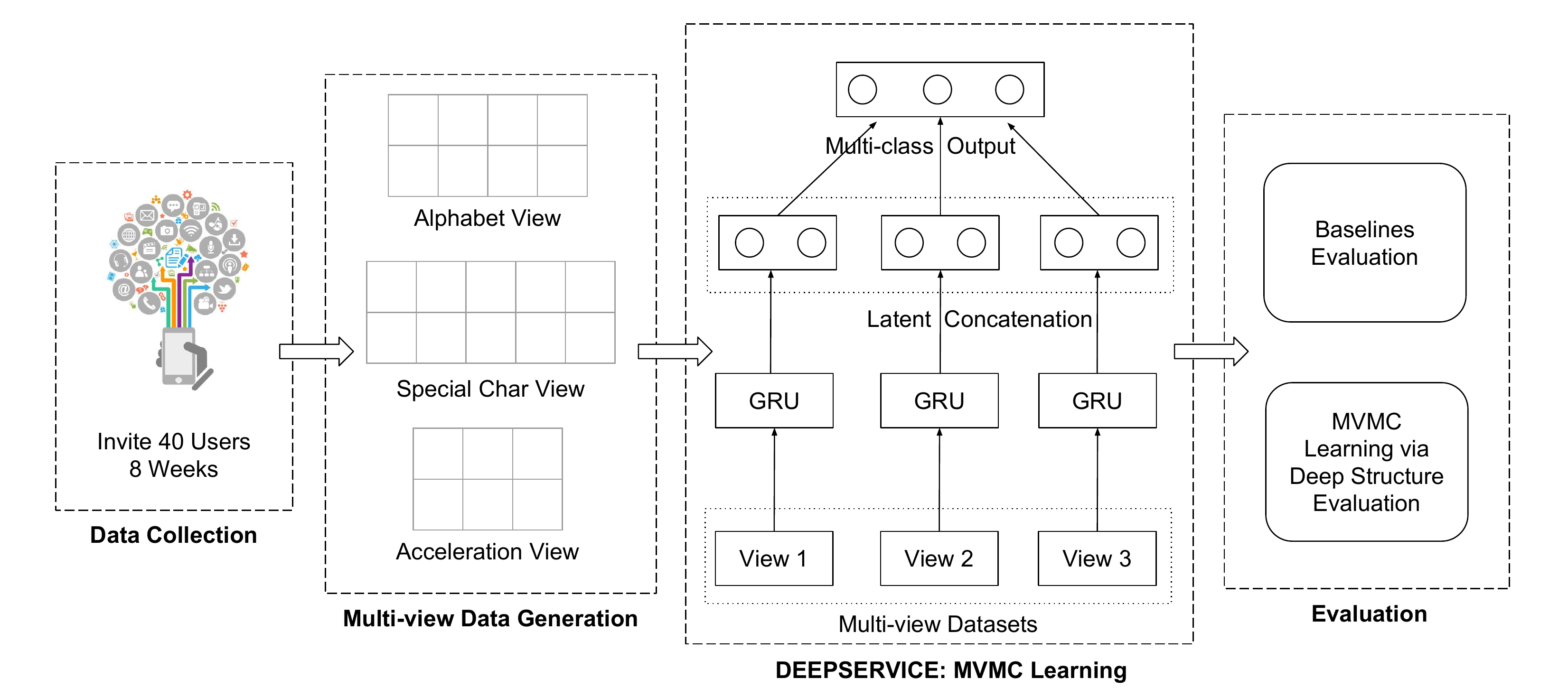}
    \caption{Framework of \textsc{DeepService}}
    \label{fig:framework}
\end{figure*}

\begin{enumerate}[leftmargin=*]
\setlength\itemsep{0pt}

\item In the first step, we collect sequential
tapping information and accelerometer information from
40 volunteers who have used our provided smartphones for 8
weeks. We retrieve such sequential data in a real-time manner.

\item In the second step, we prepare the collected information as multi-view data for the problem of user identification.

\item In the third step, we model the multi-view data via a deep structure to perform multi-class learning. 

\item In the last step, we compare the performance of the
proposed approach with the traditional machine
learning techniques for multi-class identification such
as support vector machine and random forest. This step
is discussed in Section~\ref{experiment}. 
    
\end{enumerate}

Next, we describe each of the first three steps in turn.





\subsection{Data Collection} \label{data}

First, we describe the data collection process. Our study involves 40
volunteers. The main selection criterion is based on
their prior experience with using smartphones. All
selected candidates have used smartphones for at least
11 years (some have used smartphones for 18 years). In
terms of age, the youngest participant is 30 years old
and the oldest one is 63 years old.

Each volunteer was given a same smartphone with the
custom software keyboard. Out of the 40
volunteers, we find that 26 of them (17 females and 9
males) have used the provided phones at least 20 times in 8
weeks. The data generated by these 26 volunteers is
the one we ended up using in this study. The most active
participant has used the phone 4702 times while the least
active participant has used the phone only 29 times.

\subsection{Data Processing} \label{processing}

When users type on the smartphone keyboard either
locally or on a web browser, our custom keyboard would
collect the meta-information associated with the users'
typing behaviors, including duration of a keystroke,
time since last keystroke, and distance from last
keystroke, as well as the accelerometer values along
three directions. Due to privacy concerns, the actual
characters typed by users are not collected. However,
we do collect the categorical information of each
keystroke, {\em e.g.}, alphanumeric characters, special
characters, space, backspace. Note that such
information can easily be collected from web browser
using Javascript as well.



In the data collection process, there are inevitable
missing data. For example, when the first time a user
uses the phone, the feature {\em
time\_since\_last\_key} is undefined. We replace these
missing values with 0. After the complement of missing
values, we normalize all the features to the range of
$[0,1]$.  




A typical usage of keyboard would likely result in a
session consisting more than one keystroke. For
example, a simple message such as {\em ``How are
you?''} involves sequential keystrokes as well as
multiple types of inputs (alphabets and special
characters).  In this study, one instance represents
one usage session of the phone by the user. A session
instance $s_{ij}$ represents the $j$-th session of the
$i$-th user in the data set, which consists of three
different types of sequential data. Let's denote
$s_{ij} = \{c_{ij}^{(1)},c_{ij}^{(2)},c_{ij}^{(3)}\}$
where $c_{ij}^{(1)}$ is the time series of alphabet
keystrokes, $c_{ij}^{(2)}$ is the time series of
special character keystrokes, and $c_{ij}^{(3)}$ is the
time series of accelerometer values. It is difficult to
align the sequential features in different views
because of different timestamps and sampling rates. For
example, accelerometer values are much denser than
special character keystrokes. Therefore, it is
intuitive to treat $c_{ij}^{(1)}$, $c_{ij}^{(2)}$, and
$c_{ij}^{(3)}$ as multi-view time series that together
compose the complementary information for user
identification.

\begin{figure*}[tbh!]
    \centering
    \includegraphics[width=1.0\textwidth]{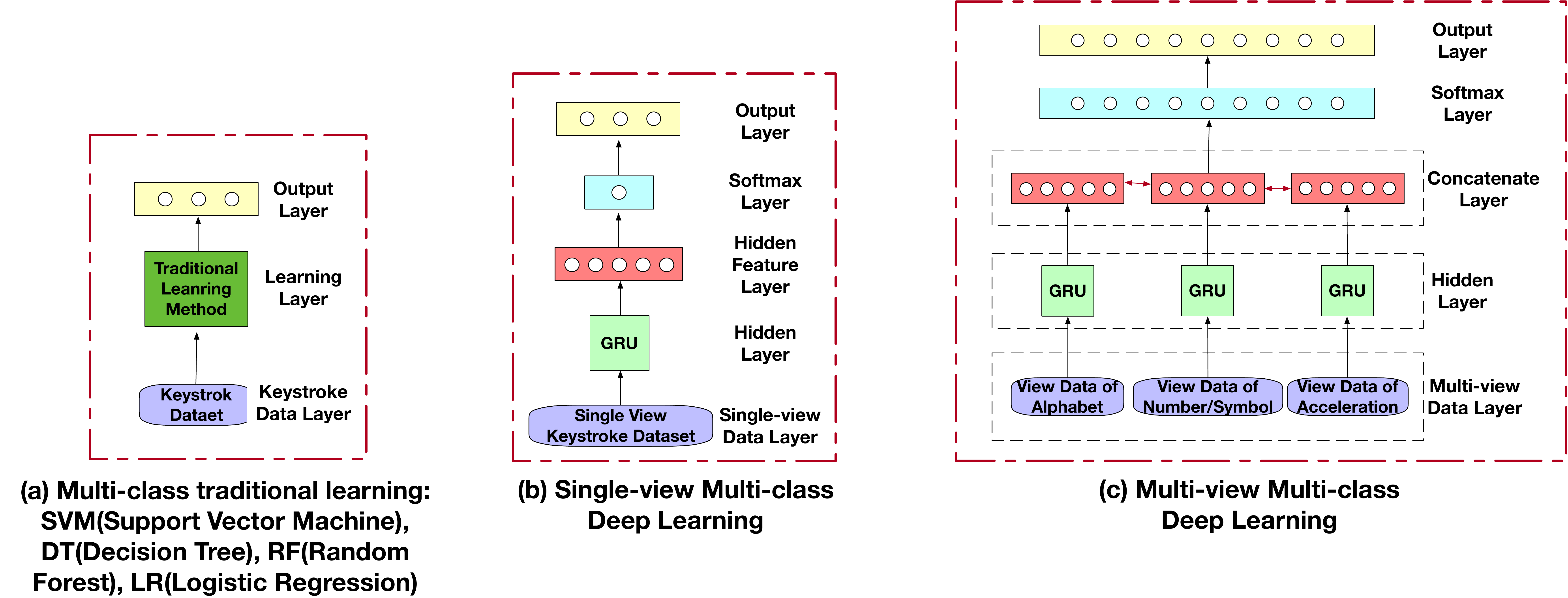}
    \caption{A comparison of different frameworks of learning models: left: (a) traditional learning methods; middle (b) single-view multi-class; (c) multi-view multi-class}
    \label{fig:mvmtl}
\end{figure*}



\subsection{Multi-view Multi-class Deep Learning (MVMC)}\label{sec:mvmc}

Now, we discuss the approach to apply deep learning for
constructing the user identification model. The
approach is based on Multi-view Multi-class (MVMC)
learning with a deep structure.

As mentioned previously, we employ three different
views. Each view $V_i$ contains different number of
features and different number of samples. In
Figure~\ref{fig:mvmtl}(a), we can use fusion method to
combine datasets of different views. However, due to
the different number of features, and the number of
records in each view of each session, it is hard to
build a single-view dataset from many other views.
Hence, instead of concatenating different views into
one view, we choose to use them separately. This is
done to avoid losing information as in the case when
multiple views are combined to create a view. One major
information that we want to preserve is the sequence of
keystrokes. By using multi-view, we are able to
maintain each view separately but then use multiple
views to make predictions~\cite{cao2017t,shao2015clustering}.
Recently, various methods have been proposed for this purpose~\cite{he2014dusk,he2017multi,he2017kernel}.

    


Before we generate multi-view multi-class learning, we
first create single-view multi-class learning as shown
in  Figure 2(b) (previously discussed in
Section~\ref{relatedwork}).  Through that model, we can
prove the multi-view can help us to improve the
performance for identification and we can determine
which view most contributes to the identification
process.  First, we separate the data set into multiple
views. In this case, we have a view of alphabets, a
view of numbers and symbols, and a view of tapping
acceleration. Then, we use GRU and Bidirectional
Recurrent Neural Network (we refer to the combination
of these two approaches as GRU-BRNN) to build the
hidden layers for each view. In the output layer, we
use softmax function to perform multiple
classifications. Finally, we can evaluate the
performance on each view.


We describe the framework that uses multi-view and
multi-class learning with deep structure in
Figure~\ref{fig:mvmtl}(c). Comparing to single-view
multi-class, multi-view multi-class is a more general
model.  After we use GRU-BRNN to build the hidden
layers for each view. We then concatenate the last
layer information from each GRU-BRNN model of each
view. We use the last concatenated layer, which
contains all information from different views, for
identification. 

As GRU extracts a latent feature representation out of
each time series, the notions of sequence length and
sampling time points are removed from the latent space.
This avoids the problem of dealing directly with the
heterogeneity of the time series from each view.  The
difference between multi-view multi-class and
single-view multi-class learning is that we use the
multiple views of the dataset and we use the latent
information of each view for prediction, which can
improve performance over using only a single-view
dataset.  We can consider single-view multi-class as a
special case of multi-view multi-class learning.





Note that in deep learning, different optimization
functions can greatly influence the training speed and
the final performance. There are several optimizers
such as RMSprop and Adam~\cite{ruder2016overview}. In
this work, we use an improved version of Adam called
Nesterov Adam (Nadam) which is a RMSprop with Nesterov
momentum. 

\section{Experiment}\label{experiment}

To examine the performance of the proposed {\sc
DeepService} on identifying mobile users.  Our
experiments were done on a large-scale real-world data
set. We also compared the results with those from
several state-of-the-art shallow machine learning
methods. In this section, we describe how we conducted
our experiments. We then present the experimental
results and analysis.

\subsection{Baselines: Keystroke-based Behavior
Biometric Methods for Continuous Identification}

In previous work on keystroke-based continuous
identification with machine learning
techniques~\cite{alghamdi2015dynamic,bo2014continuous,miluzzo2012tapprints},
Support Vector Machine, Decision Tree, and Random
Forest are widely used for continuous
identification.

\textit{Logistic Regression(LR)}: LR is a linear model with sigmoid function for
classification. It is an efficient algorithm which can handle both dense and sparse input.

\textit{Linear Support Vector Machine (LSVM)}: LSVM is widely used in many previous authorization
and identification works~\cite{alghamdi2015dynamic,bo2014continuous,miluzzo2012tapprints}.
LSVM is a linear model that finds the best hyperplane
by maximizing the margin between multiple classes. 


\textit{Random Forest/Decision Tree}: Other learning
methods such as Random Forest and Decision Tree have
not yet been adopted in many behavior biometric work
for continuous identification by keystrokes information
only. However, Decision Tree is a interpretable
classification model for binary classification. It is a
tree structure, and features form patterns are
nodes in the tree. Random Forest is an ensemble
learning method for classification that builds many
decision trees during training and combines their
outputs for the final prediction. Previous work
~\cite{sun2016sigpid} shows that tree structure methods
can work more efficiently than SVM, and can do better
binary classification comparing to SVM. We use different 
traditional learning methods on our data set as the baselines.


\subsection{Evaluation of \textsc{DeepService} Framework}




Since the number of session usage for every user is different,
we use four performance measures to evaluate unbalanced
results: Recall, Precision F1 Score(F-Measure), and Accuracy. They
are defined as:

\begin{align*}
    & Recall = \frac{TP}{TP+FN}  & F1 = \frac{2*Precision*Recall} {Precision+Recall}\\
    & Precision = \frac{TN}{TN+FP} & Accuracy = \frac{TP+TN}{TP+TN+FP+FN}
\end{align*}

Next, we report the performance of \textsc{DeepService} using
our data set.We measured
precision, recall and accuracy, f1 score(f-measure) 
for different models. Based on the
results, we can make the following conclusions.

\begin{itemize}

\item {\sc DeepService} can identify between two known
people at almost 100\% accuracy in our experiment. 
This proves that our data set contains valuable biometric information to
distinguish and identify users.

\item {\sc DeepService} can do identification with only acceleration records, even when user
is not using the keyboard. 
    
\item {\sc DeepService} is effective at  identifying a
large number of users simultaneously either locally or
on a web browser

\end{itemize}

\subsection{User Pattern Analysis}

\begin{figure*}[tbh!]
    \centering
    \includegraphics[width=0.9\textwidth]{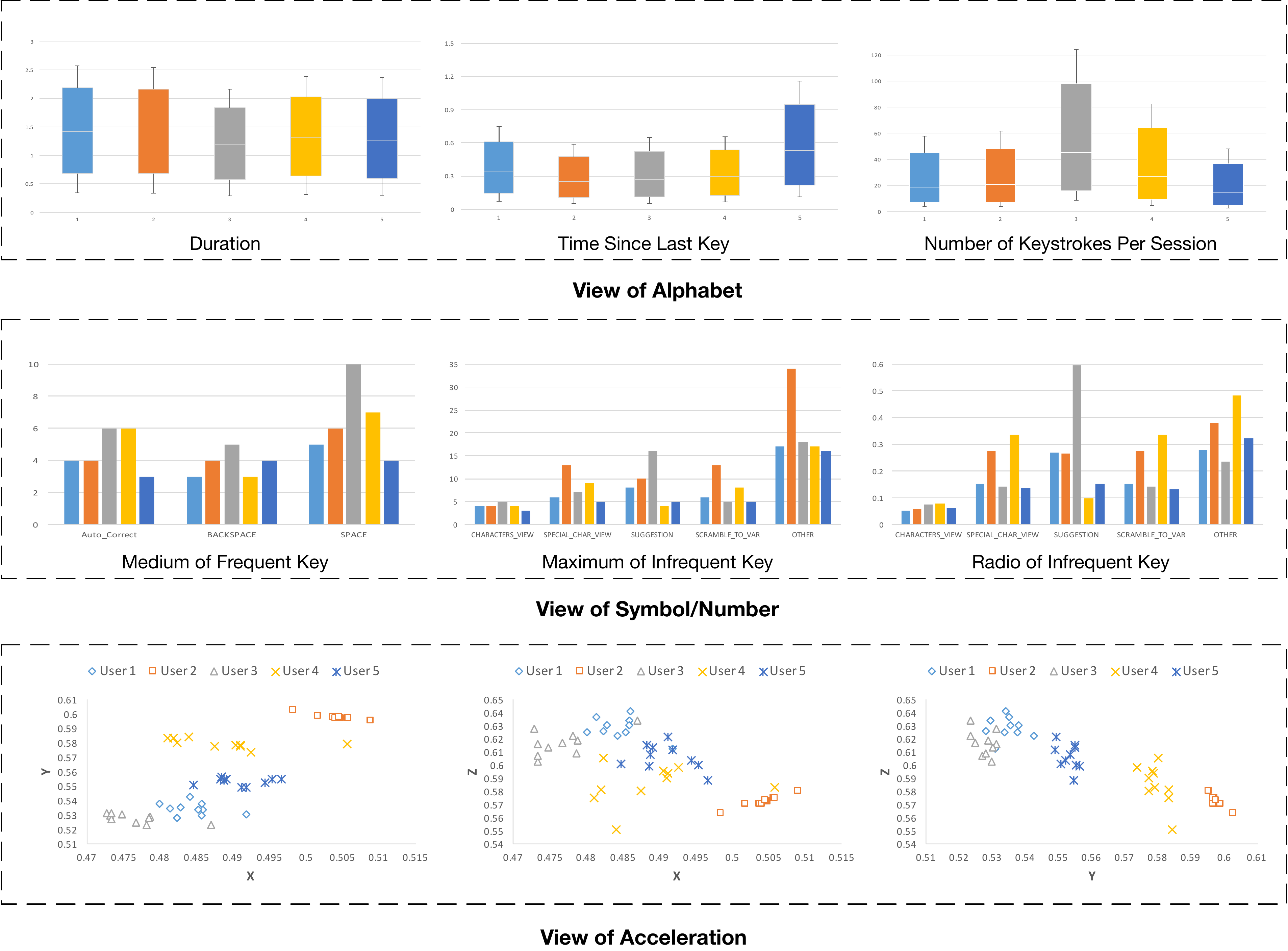}
    \caption{Multi-view Pattern Analysis of Top 5 Active User: Left is \emph{user1} and Right is \emph{user5}}
    \label{fig:user_stat}
\end{figure*}

In this section, we evaluate the feature patterns for
different users. In Figure~\ref{fig:user_stat}, we
shows the feature patterns analysis of top 5 active
users in multi-view. 

In the view of Alphabet graphs, each user tends to have
unique patterns with respect to the duration, the time
since last key, and the number of keystrokes in each
session. For example, \emph{user3} prefers to use more
keystroke in every session with quicker tapping speed
than other users. 

In the view of Symbol/Number graphs, we have 8
different features. We separate these features into two
groups: frequent keys and infrequent keys.  A frequent
key is defined as a key that is used more than twice
per session, otherwise the key is an infrequent keys.
(A user tends to use infrequent feature once per
session.).  We show the medium number of keystroke per
session of frequent keys such as auto correct,
backspace and space. We also show the range and radio
of infrequent keys per session of the top five active
users.  For example, \emph{user4} frequently uses
auto\_correct, but she infrequently uses backspace.  

In view of Acceleration graphs, we show the correlation
of different directions of acceleration. From the last
graph, we find that the top 5 active users can be well
separated, which proves acceleration can help to
identify the users well. It is also proved in our experiments.
In both user pattern analysis and experiment,
we find view of acceleration can do better identification
than other two views.




\subsection{Identifying Users}

\textsc{DeepService} can also perform continuous
identification. Before we expand to multi-class identification,
we first implement a binary-class identification based on muli-view 
deep learning which is a special case of MVMC learning identification.

\begin{figure*}[tbh!]
    \centering
    \includegraphics[width=1.0\textwidth]{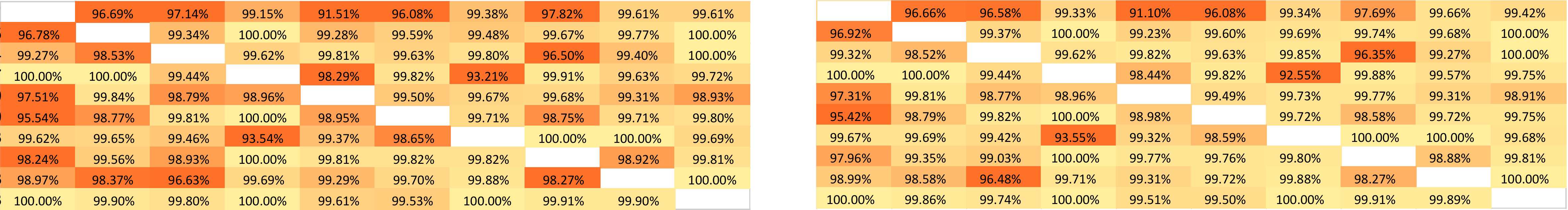}
    \caption{Heatmap of Multi-view Binary-class Identification: Left is F1 score;
    Right is Accuracy}
    \label{fig:heatmap}
\end{figure*}

In Figure~\ref{fig:heatmap}, we can see that
\textsc{DeepService} can do well identification between
any two users with 98.97\% f1 score and 99.1\% accuracy
in average. For example, private smartphone usually
would not be shared by different people. However, sometimes
the private phone could be shared between two people
, such as husband and wife.  \textsc{DeepService}
can well separate any two people in this case.

For more general scenarios, we expand binary-identification to
N active class(user) identification 
and use MVMC learning to figure out who is using the
phone either locally or on a web browser.
Table~\ref{table:MVMC} reports our results.

If we increase the total number of the users in our model,
it means we want to identify more people at the same
time. For example, if our model is used on a
home-router, it may need to identify only members of a
family (3 to 10 people) at once. If we, instead, 
want to identify people working in a
small office, we may need to identify more than 10
users. However, it is possible that a larger number of
users would degrade the average performance of user
identification.  This is due to more variation of
shared biometric patterns that introduce ambiguity into
the system. That's the main reason we want to use
multi-view data for user identification, since
different users are unlikely to share similar pattens
across all views.  

\begin{table*}[tbh!]
\caption{Results of \textsc{DeepService} and Baselines}
\centering
\label{table:MVMC}
\begin{tabular}{|l|c|c|c|c|c|c|}
\hline
                     & \multicolumn{2}{c|}{5}              & \multicolumn{2}{c|}{10}             & \multicolumn{2}{c|}{26}             \\ \hline
Method               & Accuracy         & F1               & Accuracy         & F1               & Accuracy         & F1               \\ \hline
LR             & 66.88\%          & 66.85\%          & 44.25\%          & 45.31\%          & 27.44\%          & 30.26\%          \\ \hline
SVM                  & 68.18\%          & 68.13\%          & 44.39\%          & 45.12\%          & 30.33\%          & 31.90\%          \\ \hline
Decision Tree        & 68.21\%          & 67.50\%          & 53.50\%          & 52.85\%          & 43.37\%          & 42.42\%          \\ \hline
RandomForest         & 87.59\%          & 87.42\%          & 77.05\%          & 76.59\%          & 67.87\%          & 66.31\%          \\ \hline
Deep Single View     & 82.64\%          & 82.48\%          & 78.27\%          & 78.33\%          & 61.26\%          & 63.11\%          \\ \hline
\textbf{\sc DeepService} & \textbf{93.50\%} & \textbf{93.51\%} & \textbf{87.35\%} & \textbf{87.69\%} & \textbf{82.73\%} & \textbf{83.25\%} \\ \hline
\end{tabular}
\end{table*}

\begin{figure*}[tbh!]
    \begin{minipage}{.5\textwidth}
      \centering
      \includegraphics[width=.95\linewidth]{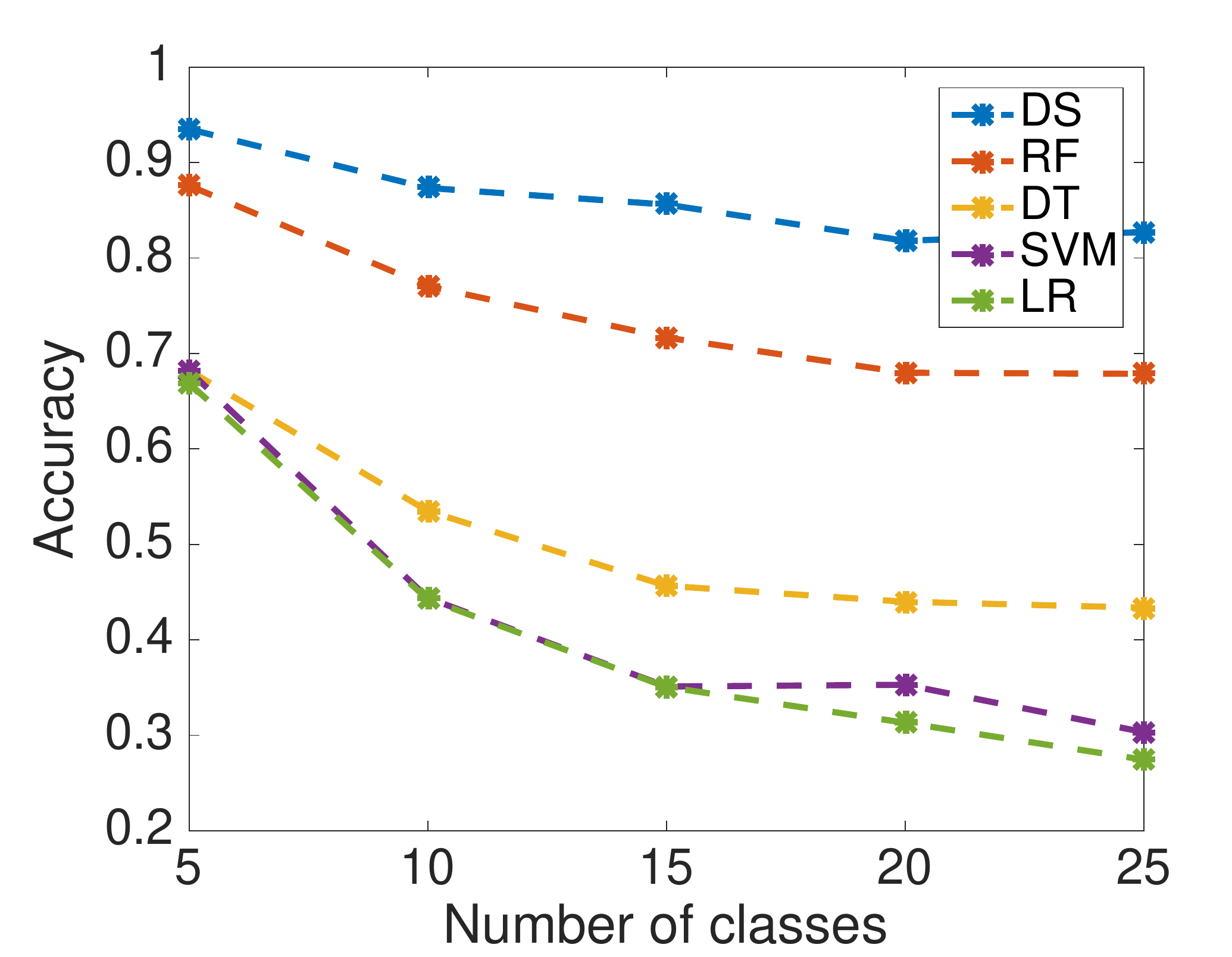}
    \end{minipage}%
    \begin{minipage}{.5\textwidth}
      \centering
      \includegraphics[width=.95\linewidth]{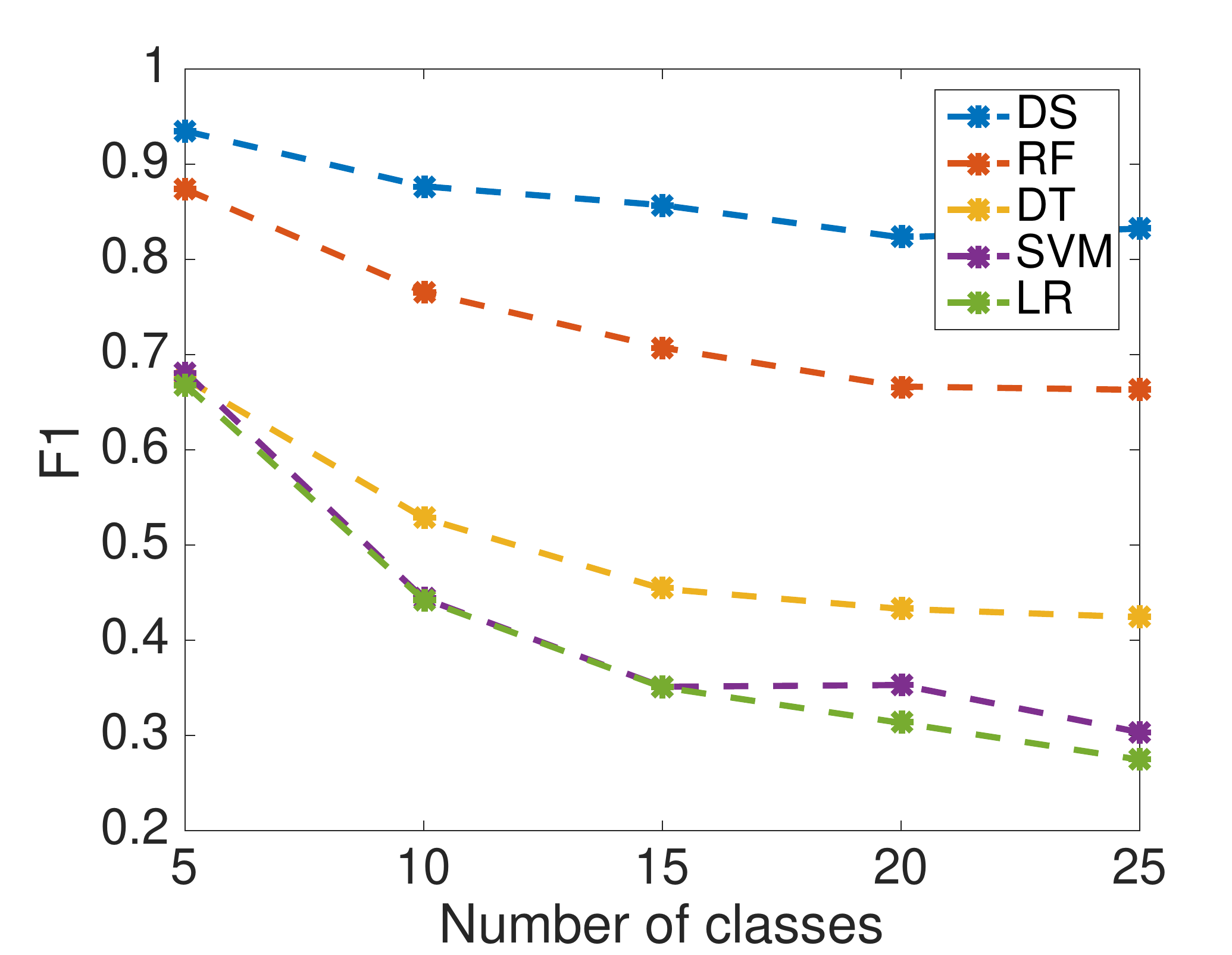}
    \end{minipage}
\caption{\label{fig:f1}Results with Incremental Number of Classes (Users)}
\end{figure*}

Table~\ref{table:MVMC} and Figure~\ref{fig:f1} report accuracy and F1 values of all learning techniques investigated in this paper. As shown, \textsc{DeepService}
can identify a user without any cookies and account
information. Instead, it simply uses the user's
sequential keystroke and accelerometer information. Our approach (DS as shown in Figure~\ref{fig:f1}) consistently outperforms other approaches listed in Table~\ref{table:MVMC}. Moreover, as we increase the number of users, the performance (accuracy and F1) degrades less than those of other approaches.

In our experiments, we also
experimented with using a single-view with deep
learning model, and found that the accelerometer view
do better identification than other two views.
However, when we used information from all three views
with MVMC learning, we achieved best performance when
compared against the results of other baseline
approaches. 

\subsection{Efficiency}

To evaluate efficiency of our system, we employ a 15"
Macbook Pro 15 with 2.5 GHz Intel Core i7 and 16 GB of
1600 MHz DDR3 memory,  and NVIDIA GeForce GT 750M with
2GB of video memory.  {\sc DeepService} is not the fastest model (decision
tree is faster), but it only takes about 0.657 ms per
session which shows its
feasibility of real-world usage.

\section{Conclusion and Future Work}

We have shown that \textsc{DeepService} can be used effectively to identify multiple users.
Even though we only use the accelerometer in this work, our results show that more
views of dataset can improve the identification performance.

In the future, we want to implement \textsc{DeepService} as a tool to help company or
government to identify their customers more accurately in the real life. The tool can be
implemented on the web or the router. Meanwhile, we will incorporate more sensors,
which can be activated from the web browser, to further increase the capability and
performance of the \textsc{DeepService}.



\section{ACKNOWLEDGEMENTS}
This work is supported in part by NSF through grants IIS-1526499,
and CNS-1626432, and NSFC 61672313.

\bibliographystyle{abbrv}
\bibliography{references}

\end{document}